# Self-assembled neuromorphic networks at self-organized criticality in Ag-hBN platform

**Authors**

*Ankit Rao[1], Sooraj Sanjay[1], Majid Ahmadi[2], Anirudh Venugopalrao[1], Navakanta Bhat,[1] Bart Kooi[2,3], Srinivasan Raghavan[1], Pavan Nukala[1]\**

**Affiliations**

[1]Centre for Nano Science and Engineering, Indian Institute of Science, Bengaluru 560012, [2]Zernike Institute for Advanced Materials, University of Groningen, Groningen, the Netherlands, [3]CogniGron center, University of Groningen, Groningen, 9747 AG, The Netherlands. Corresponding Author. * E-mail: pnukala@iisc.ac.in

**Abstract**

Networks and systems which exhibit brain-like behavior can analyze information from intrinsically noisy and unstructured data with very low power consumption. Such characteristics arise due to the critical nature and complex interconnectivity of the brain and its neuronal network. We demonstrate that a system comprising of multilayer hexagonal Boron Nitride (hBN) films contacted with Silver (Ag), that can uniquely host two different self-assembled networks, which are self-organized at criticality (SOC). This system shows bipolar resistive switching between high resistance (HRS) and low resistance states (LRS). In the HRS, Ag clusters (nodes) intercalate in the van der Waals gaps of hBN forming a network of tunnel junctions, whereas the LRS contains a network of Ag filaments. The temporal avalanche dynamics in both these states exhibit power-law scaling, long-range temporal correlation, and SOC. These networks can be tuned from one to another with voltage as a control parameter. For the first time, different neuron-like networks are realized in a single CMOS compatible, 2D materials platform.

## 1. Introduction

The efficient computational ability of the brain in tasks such as classification and pattern recognition has inspired the identification and implementation of novel computing architectures for energy efficient and parallel computing needs.[1–3] Neuromorphic hardware architectures such as crossbar memristive arrays are assembled bottom up, where individual elements of the array emulate the functionality of modular units of the nervous system i.e., a synapse or a neuron.[4–7] Using these highly organized architectures, it is difficult to capture the complexity of a biological brain, which is a self-assembled network of ~$10^{11}$ neurons and ~$10^{15}$ synapses (in humans).[8] These interconnected neuronal networks form a dynamical system, and in a state-of-rest, exhibit self-organized criticality (SOC).[9,10] In addition to neuronal cortices, SOC has also been observed in many other physical systems and phenomena such as earthquakes[11], magnetic interactions[12], ferroelastic and ferroelectric materials (Barkhausan's noise)[13,14], and charge density waves[15]. A system self-organized at critical point, when perturbed triggers successive events known as avalanches, or in other words emits 'crackling noise'.[16] The spatiotemporal dynamics of such avalanches follow power-law scaling with long range temporal correlations and scale-invariant behavior. [17–20] It has been proposed that operating at SOC allows the network to efficiently process and transport information with large scale connectivity.[21,22] A pertinent question then is whether one can engineer CMOS compatible dynamical system of self-assembled materials networks at SOC and be able to compute with these top-down neuromorphic architectures (TDNA).

In this context, self-assembled nanoscale networks comprising of nanowire networks[23] or nanoparticle complexes[24] are being recently explored. However, TDNA on CMOS compatible and large area scalable platforms are desirable for any future neuromorphic computing applications. Here, we demonstrate two different types of networks poised at SOC, both hosted in a robust 2D material platform, i.e., hBN deposited on Cu substrates with Ag as the top electrode. The choice of high quality hBN as an insulator in conjunction with fast diffusing Ag as the top electrode, allows us to uniquely stabilize, a) percolative tunnel junction network of Ag intercalates and b) Ag filamentary network. The 2D material platform also allows for heterogenous incorporation across different device architectures.[25]

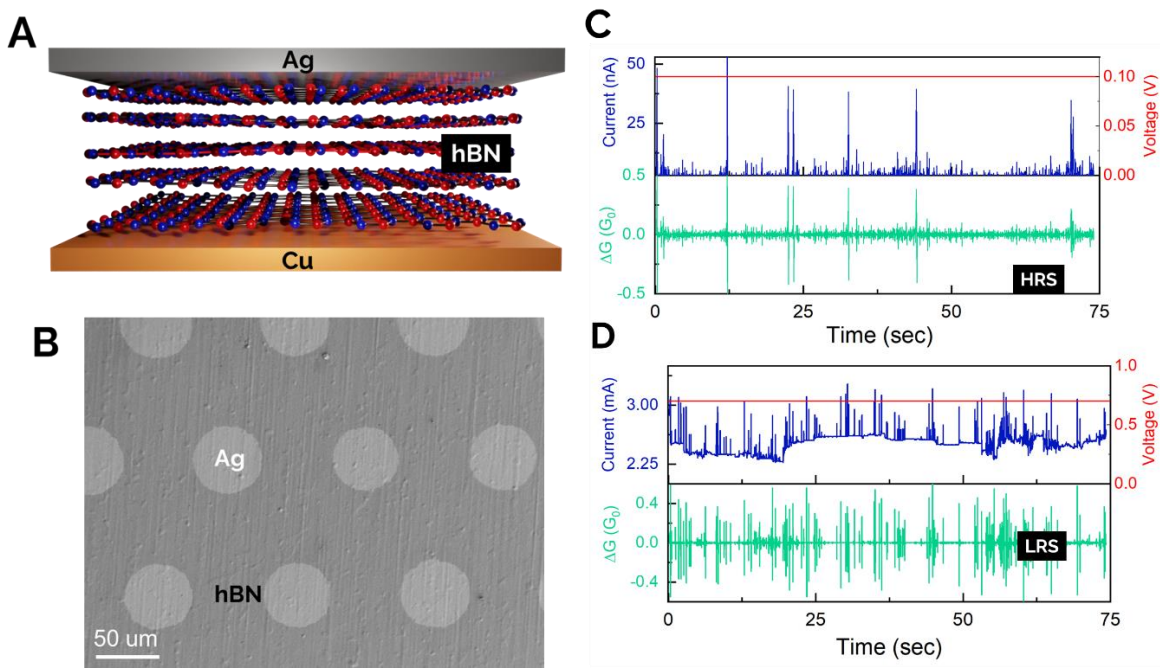

**Figure 1. Avalanche Dynamics in Ag-hBN system.** A) Schematic of Ag-hBN-Cu (MIM) device. Voltage is applied to Ag, and Cu substrate is ground. B) SEM image of the top view of the devices. C) Representative current data (blue) when sampled at a constant voltage (red) as a function of time in HRS state, showing stochastic spiking events and the differential conductance data representing the event size ΔG (red) expressed in quanta of mean conductance, $G_0$. The current levels observed are in nA. D) Representative current and differential conductance data in the LRS state. The current levels observed are in mA.

## 2. Results

### 2.1. Dual state switching of Ag networks

The growth of large area (upto 6 sq inches) multilayer hBN (~10nm thick) on Cu was carried out by chemical vapor deposition process as described in the methods section which has been previously reported.[26] Metal-insulator-metal devices arrays (25x25) were fabricated with Ag (50 um) as the top electrode, hBN as the insulator and Cu substrate as the bottom electrode (Figure 1A, 1B and S1A). The measurements were performed on 3 different samples with 10 devices measured on each sample, all of which showed similar behavior. Cu was ground, and voltage was applied to Ag electrode. I-V sweeps revealed bipolar and bistable resistive switching (Figure S1B), after an initial forming step (Figure S1C). In all these devices HRS to LRS transition occurs at $V_{set}$= 0.9V (±0.15V), and LRS to HRS transition at $V_{reset}$= -0.6V (±0.1V). To characterize the spatio-temporal avalanches in both HRS and LRS, we performed voltage sampling measurements i.e., measured instantaneous current, by applying a constant voltage bias. Representative device stabilized in HRS and LRS was tested at small perturbing voltages of 0.1 V and 0.6 V respectively. Current response in HRS (Figure 1C) shows irregular spikes of various amplitudes, ranging from 5 nA to 50 nA. Response in LRS also shows irregular spikes of a much greater amplitude in the range of 2.2 mA to 3 mA (Figure 1D). The statistical behavior and correlations of these spikes can be understood through order parameters that characterize them, such as change in conductance, $\Delta G$ ($G_o$) (Figure 1C and 1D), inter event/spike interval (IEI) and autocorrelation function of IEI (ACF).

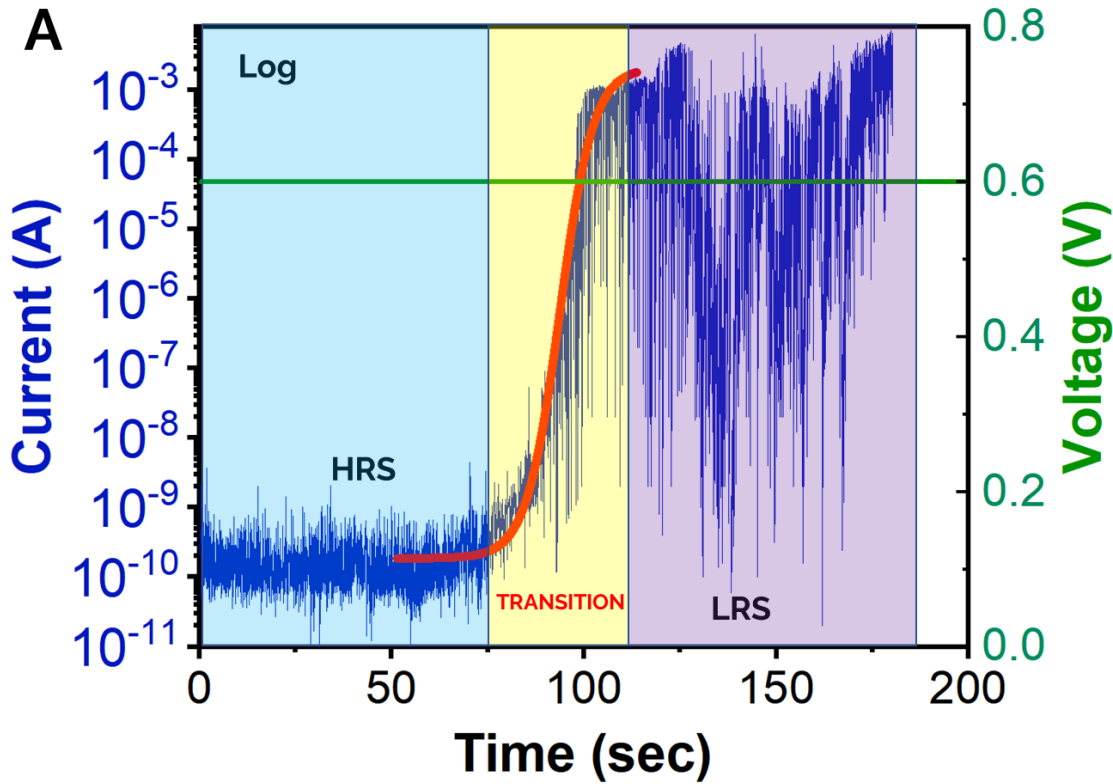

**Figure 2. Diffusion based State Transition in the Ag-hBN system.** Log scale plot of the behavior observed. Different colors indicate different stages of device, HRS (blue), Transition from HRS to LRS (Yellow), LRS (Purple). JMAK model fit for the transition is shown in red solid line.

At voltages close to, yet smaller than the set voltage, devices in HRS undergo a phase transition to LRS in measurable time scales. For e.g., a representative device prepared in HRS and held at 0.6 V ($V_{set}$= 0.9V) exhibits nano ampere current fluctuations upto 70 sec in HRS, followed by a transition to LRS and milliampere current fluctuations in LRS (Figure 2, linear scale Figure S1D). The average current fits well to John-Avrami-Mehl-Kolmogorov kinetics in the transition region[27] as per Eq (1),

$$i_T = at^{n-1}e^{(-bt^n)} \qquad (1)$$

where $a$ is a constant, $b$ is a composite parameter which includes the kinetic rates of nucleation and growth, and the parameter $n$ is the Avrami exponent, where $n = d + 1$, $d$ being the dimension of the system. From our fits, we estimate n~2 (Figure S2A), likely suggesting a 1D growth process consistent with the formation of Ag filaments through long-range diffusion.[28] Subsequent large amplitude current fluctuations in LRS suggest sporadic events of breakage and reformation of the conducting filaments.

The spiking rates observed in HRS and LRS are dependent on the sampling rates. For the sampling rates of 10 ms, the measured spiking rates for the representative device are in the range of 15-20 spikes per second in HRS and 25-30 spikes per second in LRS. Furthermore, spiking rates are regulated in both HRS and LRS with respect to the voltage applied (0.05V to 0.5V; Figure S2C). Such rate coding allows for encoding the device with voltage dependent spike rate information for stochastic device applications.[29] Finally, for the devices prepared in LRS, upon the application of negative voltage close to the RESET voltage (-0.6V), there is a discrete and abrupt transition to HRS state as seen in Figure S2B, followed by spiking behavior in the HRS. Notice that under negative bias there is no spiking behavior in LRS.

**2.2 Long Range Temporal Correlation and Avalanche dynamics**

To better envisage the avalanche dynamics, the current spiking pattern is binarized by classifying those spikes which exceeded a threshold limit as events and rest as not. The top panel event train in Figure 3A (HRS) and 3E (LRS) contains data acquired for 100 s, while the temporal scales are magnified 5 and 20 times for the subsequent panels. The event train patterns in all the panels look qualitatively self-similar and are independent of the sampling rate. The switching activity in Figure 1C and 1D (binarized subsections in Figure 3A and 3E respectively) exhibits event bursts with varying range of ΔG values. These bursts indicate a consecutive correlated activity distributed across the network, rather than isolated local events. To quantify the correlation and self-similarity behavior, we investigate the distribution of event sizes (ΔG), inter-event interval (IEI) and the autocorrelation function of the IEIs (ACF).

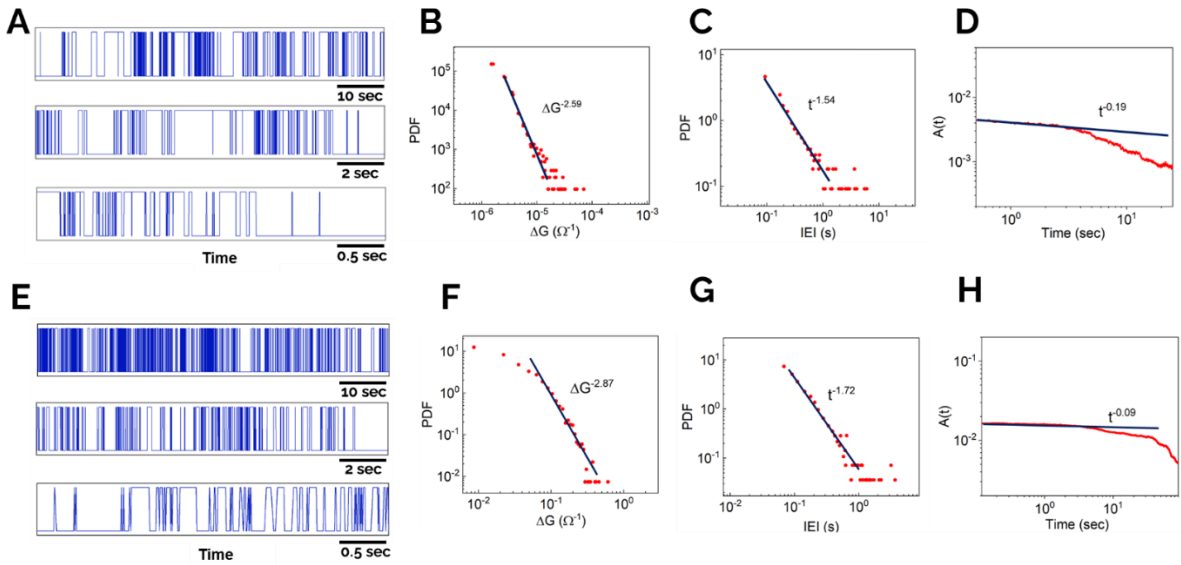

**Figure 3. Long Range Temporal Correlation (LRTC) and Power Law Scaling.** The distribution plots shown in (A to D) are for the device in HRS state and (E to H) in LRS state. (A and E) The binarized patterns of the switching activity in the Ag-hBN-Cu device, plotted for varying time scales. These patterns exhibit self-similar temporal behavior. (B and F) The probability density of change in the conductance of the device (ΔG), plotted for the avalanche patterns in HRS (B) and LRS (F) states. (C and G) The power law distributions for the Inter event intervals (IEIs) are demonstrated as the probability distribution functions (PDF). (D and H) Autocorrelation function of IEIs in both HRS and LRS, both of which show a slow power law decay.

Figure 3B and 3F shows the probability distribution functions (PDF) of event sizes in HRS and LRS. The P(ΔG) in both states follows a power law distribution spanning 3 orders of magnitude with exponents of ~2.5±0.1 (HRS) and ~2.8±0.1 (LRS). The PDF of the inter-event intervals (IEIs) (Figure 3C and 3G) also follow a power-law distribution with exponent, γ in the function, (P(t) ~ $t^{-\gamma}$) ~1.54±0.05 in HRS and 1.72±0.05 in LRS. This is a quantitative proof of self-similar, scale free nature of the events (qualitatively also observed in Figure 1A and 1E). Self-similarity is a signature of temporal correlations which are observed in electrophysiological signals from brain cortex and are an outcome of fractal correlations.[30] The γ in HRS is similar to the values obtained in nanoparticles systems exhibiting percolating tunnel junction networks[19] and the γ in LRS is similar to the values in nanowire networks[23]. These values are also in similar range as those observed in neuronal cortices.[31,32] The ACFs in both HRS and LRS (Figure 3D and 3H) exhibit a slow power law decay over several decades. Another order parameter of interest to understand the correlation between successive events is inter-spike interval ($ISI_n$), defined as the time between rising edges of consecutive spikes. Figure S4A (in HRS) and Figure S4C (in LRS) show a tight scatter of $ISI_n$ and $ISI_{n+1}$ about the mean ISI. The observed ACF and the $ISI_n$ behavior is a result of long-range temporal correlations (LRTC) among events. Finally, as is the case with any correlated noise, the event power spectrum follows a power law with frequency as $1/f^\beta$. β is 0.91 in HRS (Figure S4B) and 0.82 in LRS (Figure S4D) in our representative device.

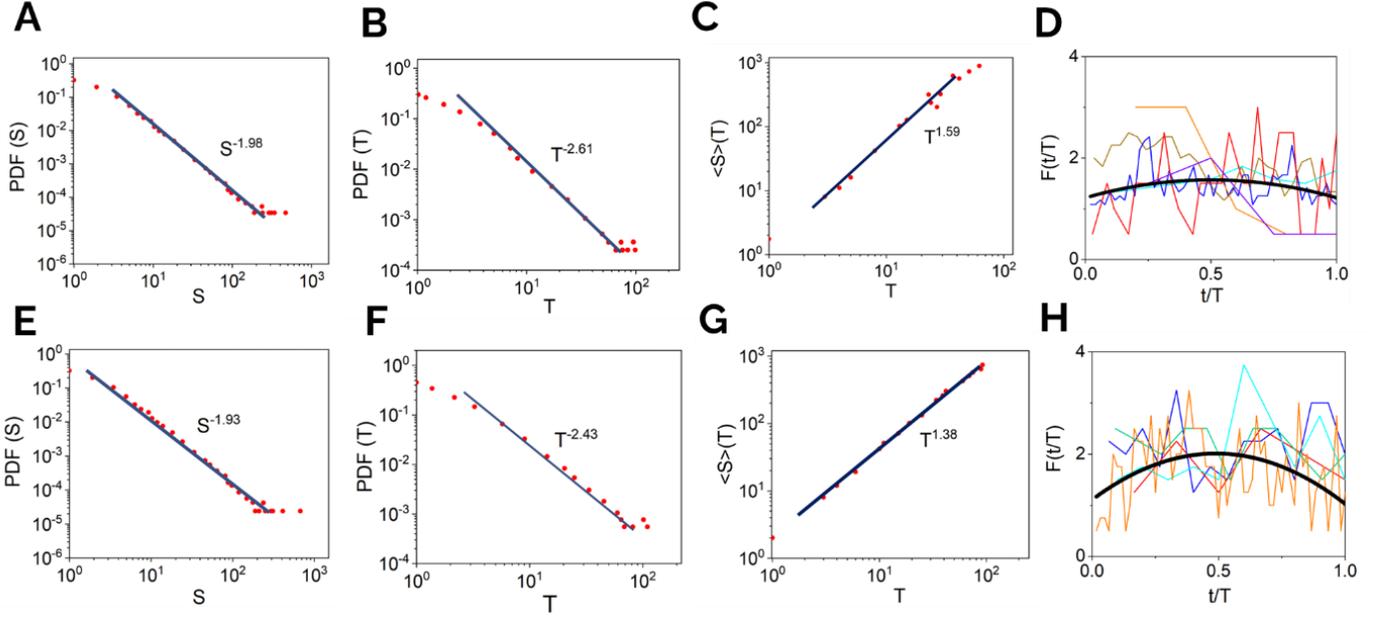

**Figure 4. Self-organized criticality.** The distribution plots in A to D are for the HRS whereas plots E to H are for LRS which exhibit self-organized behavior across numerous temporal scales. (A and E) The probability distribution function for the size of an avalanche P(S) in both cases follows a power law with similar exponent values of τ ~1.95. (B and F) Distribution of duration of avalanche P(T) also follows power laws in both states with exponent values of α~2.6 (HRS) and α~2.45 (LRS) with linear binning. (C and G) The average avalanche size per unit time bin <S>(T) also exhibits power law scaling with exponent 1/συz ~ 1.5 (HRS) and 1/συz ~ 1.4 (LRS). (D and H) The mean avalanche shapes for avalanches of various Ts (different colors) collapse onto a universal parabolic scaling function (black solid line). The exponent values obtained fulfill the crackling relationship (Eq 5) which establishes the existence of critical avalanche dynamics. In (D) avalanche duration, T=4 (Purple), 4 (Teal), 5 (Orange), 12 (Brown), 15 (Red), 24 (Blue)], and in (H) T= 4 (Red). 5 (Green), 6 (Teal), 10 (Green), 32 (Orange)]

### 2.3 Self-organized criticality observation

An avalanche is defined as the sequence of successive event occurrences between two empty time bins. The size of an avalanche (S) in such an occurrence is the total number of events in the avalanche, time of avalanche (T) is the number of time bins and the (<S>(T)) is defined as the average size of the avalanche of a duration T. In a critical system, the probability distribution functions of S, T and <S>(T) follow power-law scaling as described in Eq (2), Eq (3) and Eq (4) respectively.[33] The distributions are defined as follows:

$$P(S) \sim S^{-\tau} \quad (2)$$

$$P(T) \sim T^{-\alpha} \quad (3)$$

$$<S>(T) \sim T^{1/\sigma \upsilon z} \quad (4)$$

A signature of avalanche activity which exhibits SOC is when the power law exponents obtained agree with the crackling relationship as shown in Eq (5).[34]

$$\frac{1}{\sigma \upsilon z} = \frac{\alpha - 1}{\tau - 1} \quad (5)$$

Eq (5) distinguishes a system poised at SOC from other random walk processes that also yield power laws and exponents. In both HRS and LRS as shown in Figure 4, the switching activity exhibits power law dependence for P(S), P(T) and <S>(T) with power law exponents τ, α and 1/σνz respectively. In HRS, τ ≈ 1.96±0.05 (Figure 4A) and α≈2.61±0.05 (Figure 4B), which gives 1/σνz ≈ 1.67±0.1 from Eq (5). This is in close agreement with 1/σνz ≈ 1.59±0.05 independently estimated from <S>(T) (Figure 4C). Similar agreement is observed in LRS case, with τ ≈ 1.93±0.05 (Figure 4F) and α≈2.43±0.05 (Figure 4G), which gives 1/σνz ≈ 1.52±0.1 from Eq (5), which is also in agreement with 1/σνz ≈ 1.38±0.05 obtained from <S>(T) (Figure 4H). Table S1 summarizes the exponents observed in various order parameters in both HRS and LRS and compares it with the corresponding exponents in neuronal cortices and other systems. This validation of the crackling noise relationship proves that our system is poised at SOC in both HRS and LRS.

Another crucial feature of a system at criticality is that the average shape of the avalanches exhibits universal scaling or shape collapse.[34] The universal scaled shape is obtained by normalizing the duration of avalanche with respect to the largest avalanche duration, T, and size of the avalanche with respect to the average size of the avalanche of duration as described in Eq (6) and (7). [35]

$$s(t,T) \propto T^{\left(\frac{1}{\sigma v z}-1\right)} F\left(\frac{t}{T}\right) \quad (6)$$

$$<S>(T) = \int_0^T s(t,T)dt \quad (7)$$

Here, $s(t,T)$ is the mean number of spikes (s) at a time t in an avalanche with duration T and $F\left(\frac{t}{T}\right)$ is the shape of the avalanche when plotted for a scaled duration (t/T). We perform this shape collapse analysis for all the avalanches of different times T and observe universal parabolic scaling that has also been reported in neuronal avalanches.[35,36] The shape collapse analysis yields another independent estimate of the 1/σνz which is 1.5±0.05 for HRS (Figure 4D) and 1.4±0.05 for LRS (Figure 4I). These values are similar to those obtained from Eq (5) and provides another independent verification of 1/σνz. The exponents 1/σνz, α and τ do not show any significant variation with voltage in both HRS and LRS (Figure S5). The close agreement of 1/σνz obtained independently from the crackling relationship estimates, shape collapse analysis and Eq (5) provides robust evidence of SOC in HRS and LRS in the Ag-hBN system.

The network structure of HRS and LRS can be understood from the cross-sectional TEM analysis shown in Figure 5. Cross-sectional lamellae of representative devices prepared in pristine state, HRS (after 10 cycles), and LRS (after 20 cycles) were made using focused ion beam technique. Figure 5A shows HRTEM image of pristine device with high quality crystalline two-dimensional layers of hBN grown on Cu substrate and contacted with Ag. In HRS, the HAADF STEM (Figure 5D) and EDS analysis (Figure 5F) of the selected region shows dispersion of Ag intercalates inside the hBN matrix (also see Figure S6A-E). The HRTEM data and analysis further shows lattice expansion from 0.35 nm in pristine state (Figure 5B) to 0.39 nm in HRS (Figure 5C), which is a consequence of intercalation of Ag inside the h-BN matrix. Similar expansion resulting from intercalation has been observed in other 2D systems.[37] HAADF STEM and EDS analysis of the device in the LRS state shows formation of multiple Ag filaments across the h-BN film (filamentary networks, Figure 5F, 5G and Figure S6 I-J).

## 3. Discussion

A simple model proposed hereunder allows us to explain all the above results in our system. HRS and LRS form physically two different types of networks, both poised at SOC. It is important to point out that such a unique behavior occurs due to a combination of fast diffusing species in Ag (top electrode) and CVD grown high quality robust hBN matrix. We confirmed the uniqueness of this structure by contacting hBN with standard top contacts used in memristive devices such as Ni and Au, both of which did not exhibit any spiking activity (Figure S7A and S7B).

Pristine hBN electroforms to an LRS beyond a threshold voltage through the formation of Ag filaments as shown in Figure S1C. Ag is known to be highly diffusive in nature and the ionization potential for Ag to $Ag^+$ transition was reported to be 2-2.5 eV lower than other metals such as Au, Cu, and Ni.[38] This allows Ag to ionize at relatively lower electrochemical potentials and migrate owing to its comparatively small ionic size through the hBN matrix to the Cu electrode.[39] Here nanoclusters are nucleated which grow in the form of filaments. Following electroforming, the device is cycled between LRS and HRS several times.

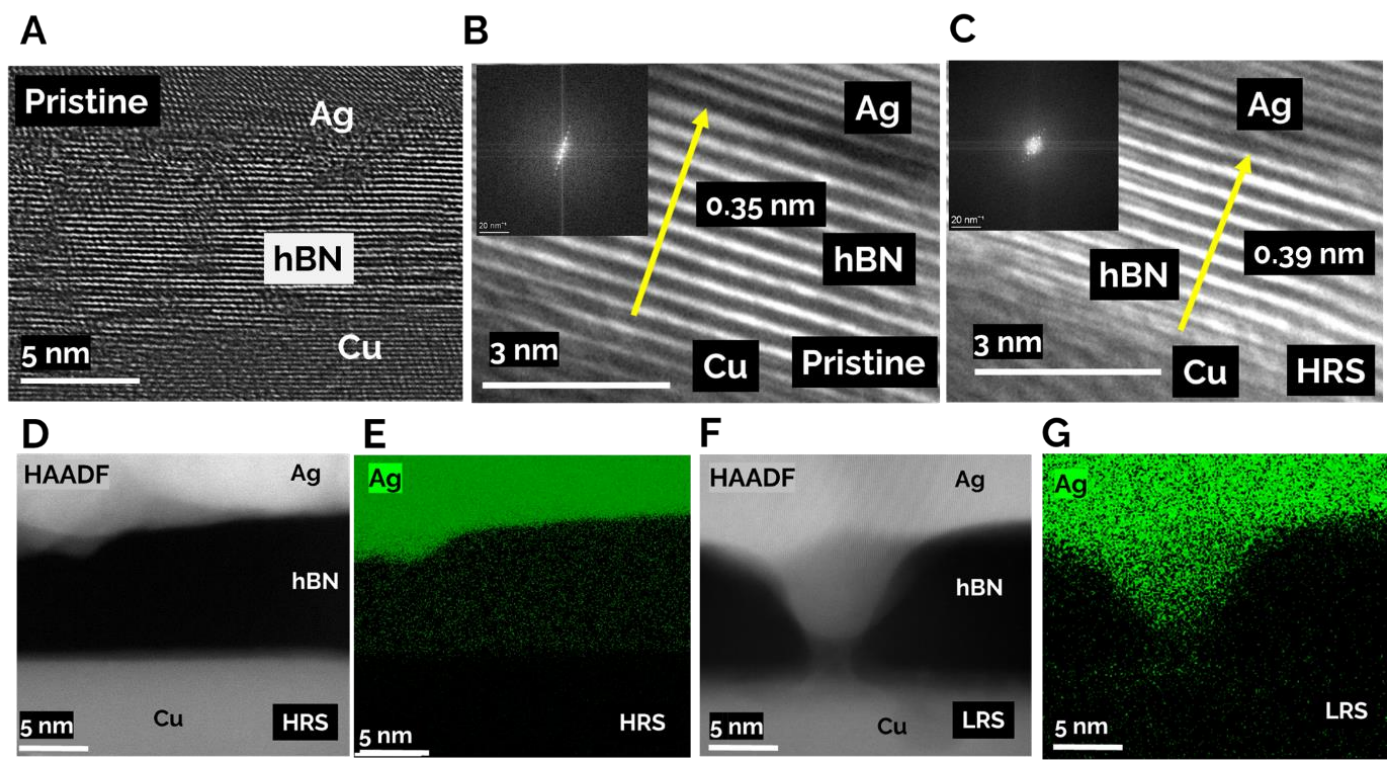

**Figure 5. Transmission Electron Microscopy Analysis of Ag-hBN system.** A) Cross-sectional high-resolution transmission electron microscopy (HRTEM) image of the as grown (pristine) hBN on Cu contacted with Ag. B) HR-TEM of the hBN stack in pristine state exhibiting interplanar distance of 0.35 nm. (Inset) FFT of the hBN region showing layered structure. C) HR-TEM of the region after multiple cycles and spiking in HRS state, the resulting interplanar distance is 0.39 nm. (Inset) FFT of the hBN region exhibiting ring patterns due to presence of Ag particles along with the layered structure pattern. D) HAADF-STEM image in the HRS state of the Ag-hBN-Cu stack. E) EDS map of the area reveals presence of Ag clusters further confirming the ionic movement of Ag inside the hBN matrix. F) HAADF-STEM image in the LRS state shows formation of the Ag filament due to large scale diffusion of Ag under continual cycling to the LRS state. G) EDS map of Ag in the area shown in (F).

HRS, as evidenced from TEM data, contains Ag nanoclusters which intercalate into the van der Waals gaps of the hBN matrix (schematic in Figure 6A). The high crystalline quality of hBN planes (low defect density), means that while intralayer diffusion of Ag can be fast, the interlayer diffusion, mediated by defects such as point defects, stacking faults and grain boundaries is a slower process. This allows for intralayer clustering, and formation of a stable network of Ag intercalates (nodes). Intercalation of species such as Ag is well reported in other 2D systems too.[37,40,41]

The conduction in the HRS state (2 orders higher than the pristine state) is predominantly mediated by electron tunneling across these nodes, with hBN acting as tunnel barrier between interlayer nodes, and vacuum as the barrier between intralayer nodes. In such a self-assembled tunnel network of Ag nodes, application of a small perturbing voltage creates fluctuations in the clusters/nodes, dynamically changing the spacing between them, thus giving rise to correlated nanoampere fluctuations in tunnel current (Figure 1C). These events are polarity insensitive and happen both in positive and negative biasing voltages (Figure 1C, Figure S2B). This network self-organizes to criticality, characterized by scale free avalanches analogous to sand pile at a critical slope, and corresponding avalanches created upon a small perturbation to maintain the critical slope.

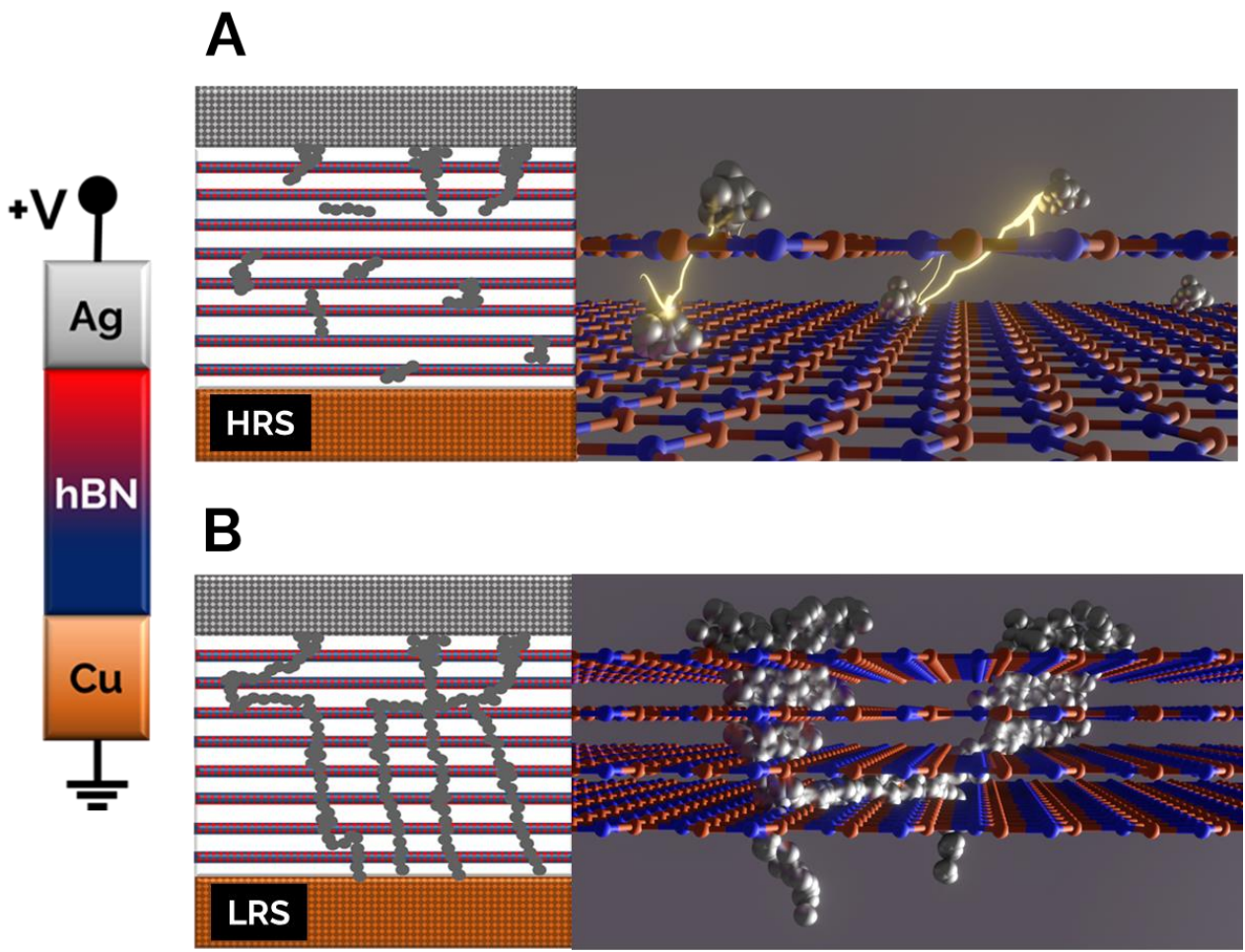

**Fig. 6. Avalanche Mechanism** A) The Ag acts as the top electrode, which is biased, and the Cu substrate acts as the bottom electrode which is the common terminal. The device schematic when cycled and rested in HRS state showing the Ag (grey) gets dispersed in the hBN matrix. Schematic of the tunneling current between the Ag clusters which are formed between the hBN layers. B) The device schematic when cycled and rested in LRS state showing the formation of filamentary networks. 3D representation of the filamentary networks formed within the hBN matrix.

The interlayer migration of Ag, which is a slower process at low voltages, dictates the process of filament formation, and thus the transition between HRS and LRS. The kinetics of this transition can be non-linearly enhanced by increasing the voltage, a phenomenon often referred to as voltage-time dilemma in resistive memory.[42] Thus, by operating our representative devices at voltages of 0.6 V ($V_{set}$=0.9 V), an HRS to LRS transition is recorded over the background of current fluctuations as shown in Figure 2A. The kinetics fit JMAK model, with the dimension of the phase transition ~1, strongly supporting filament formation. LRS is thus a network of multiple Ag filaments which are further directly imaged using HAADF-STEM (Figure 5G, Figure S6 I, J). Application of a negative reset voltage beyond a threshold, results in electrochemical dissolution of the filaments, giving rise to the Ag intercalated network structure of the HRS, and bipolar memristive switching.

The milliampere current fluctuations in LRS occur when a small positive voltage is applied on the system. This can be explained as a competing series of filament breaking and mending events, with Joule heating and corresponding random surface ion diffusion causing the filaments to break, and ionic migration and redox reactions at the electrodes (electrochemistry) supporting their mending. These events also show long range temporal correlations, and scale free nature. However, when a negative bias is applied in such filamentary networks (LRS), both electrochemistry and Joule heating cause the filaments to break, and thus our devices did not show any spiking behavior (Figure S2B). We also observe diffusion based self-stabilizing behavior under application of long-range pulsed stimulus (Figure S3).

Both the tunnel networks (HRS) and the filamentary networks (LRS) are poised at self-organized criticality, showing a universality behavior, parametrized by critical exponents of various order parameters. In the HRS the values of $\tau$, $\alpha$, $1/\sigma\nu z$ correspond to the values discovered in nanoparticle percolative tunnel networks as seen in Table S1, suggesting not only that these networks belong to the same universality class, but also that the underlying dictating mechanism is the formation and dissociation of percolative paths of tunneling currents. In LRS, however $\tau$, $\alpha$ and $1/\sigma\nu z$ show marginally different values, probably indicating that this could belong to a different universality class. The exponents obtained are similar in values to those in the nanowire networks which are filamentary in nature. The critical exponents observed in both these classes are different from the well-known universality classes such as mean-field, propagating front, short-range models in various dimensions, established, for e.g., in studies on magnetic Barkhausan's noise or charge density waves.[13–15] To comment any further requires rigorous modeling of these types of networks. Our work opens a fundamental problem in physics of identifying various universality classes of different types of neuromorphic networks. It is our belief that correlating the universality classes with their computational performance will be a crucial step in the development of top-down neuromorphic architectures.

## 4. Conclusion

In this work, we demonstrate a unique CMOS friendly platform which hosts neuronal-like networks of two different universality classes (percolative tunnel junction networks and filamentary networks), both poised at SOC. These networks can be transformed between each other through electrical biasing, aided by field dependent migration kinetics of Ag in the hBN matrix. Such observations are also first of their kind in any 2D material platform. The distinct advantage of our large area CVD grown hBN platform is that it can be transferred to any substrate, including Si, allowing for the implementation of integrated systems on large-area devices. Furthermore, such 2D platforms allows for engineering new structures via heterogenous layer stacking for future applications in neuromorphic computing. As an outlook, we envision arrays of these devices (multielectrode arrays) which can emulate spatiotemporal dynamics similar to those observed in neuronal cortices and can mimic brain-like operations. To compute using these platforms, approaches such as reservoir computing will be a good starting point.[43,44]

## 5. Materials and Methods

### 5.1 Growth and Device Fabrication

The growth process of multilayer hBN on Cu was carried out using BH3NH3 (Sigma Aldrich, 97%, 682098) as the precursor in a custom built hot-wall CVD reactor. We employed a temperature and pressure-controlled vaporizer setup to control the flux of the precursor entering the reactor to grow multilayer hBN. A 1 cm x 6 cm Cu foil (Alfa Aesar, 99.8%, 46986) of 25 μm thickness was normally used. The foil was washed in dilute hydrochloric acid for 5 min to eliminate the oxide layer on the surface along with a deionized (DI) water rinse. Successive cleaning steps were conducted in acetone and isopropyl alcohol for 5 minutes in each solvent for eliminating the organic contaminations on the surface. The foil was then immediately inserted into the reactor compartment at a permanent location. The compartment was then pumped to the base pressure flushed with 200 sccm of $H_2$ for 10 min and then ramped up to the growth temperature which was usually between 1025°C and 1050°C. The details of the growth process can be found in the previous work. Ag contacts (50 um) were thermally evaporated on the as-grown hBN-Cu with the aid of stencil masks. The electrical measurements for the metal-insulator-metal stack of Ag-hBN-Cu were carried out as seen in Figure 1B.

### 5.2 Measurements

DC probe station (Agilent Device Analyzer B1500A) was used for electrical characterization. Voltage was applied on the top electrode (Ag) and the bottom contact of Cu was held at ground potential. The primary method of measurements adopted is application of constant DC voltages within the I-V-t (current-voltage-time) sampling module. This allows for study of current reconfigurations of the local switches in the device leading to spikes and thus, the analysis of avalanche dynamics. These measurements are carried out at multiple sampling rates, such 100 us, 1 ms and primarily at 10 ms sampling interval. Larger sampling interval allows for large data set and extended time sampling which helps to circumvent substantial breaks in the power law distributions. The data obtained for all the sampling intervals demonstrate similar data in quality and quantity, with similar power law exponents upon analysis. Thus, the results obtained are indifferent to the sampling rate. In this work, we show data from ten devices on a single sample, the data and analysis of further 3 samples and 30 devices are consistent with the DC and pulsed measurements performed. Scanning Electron Microscopy was performed using Zeiss Ultra 55 FESEM, Zeiss. Transmission Electron Microscopy was performed Dual-aberration corrected Thermofisher Themis microscopes operated at 300kV. The cross-sectional TEM of the sample was prepared using focused ion beam (FIB) Helios G4-UX.

### 5.3 Analysis

The analysis of the electrical data has been carried out by methods similar to those used to analyze multi electrode array measurements in neural cortices in neuroscience and biological studies. The data analysis routines are borrowed from ref [35], with minor modifications. Thresholding method is used to define changes in the conductance value as per the current signal when exceeds a threshold value. The threshold value is calculated according to the mean range analysis so as to not alter the significance of IEI distributions and autocorrelation function. The maximum likelihood estimation (MLE) approach is incorporated to estimate the power-law exponents in the spiking distributions. The MLE fitting algorithm estimates the best fit for the distribution. The MLE estimators for the distributions are calculated according to the Kolmogorov Smirnov (KS) test which identifies the variance from the power-law fits. We accept the fit for a KS-statistic (p) of the data that was less than the KS-statistic found for at least 20% of the power-law models ($p_t$) (i.e., $p \geq p_t = 0.2$). For the autocorrelation function, the standard linear regression method is used to obtain the exponents. For shape collapse analysis, a polynomial fit using least squares fitting is used to obtain the curvature scaling parameter which minimizes the variance.

**Acknowledgements**

This work was partly carried out at Micro and Nano Characterization Facility (MNCF), and National Nanofabrication Center (NNfC) located at CeNSE, IISc Bengaluru, funded by NPMAS-DRDO and MCIT, MeitY, Government of India; and benefitted from all the help and support from the staff. P.N. acknowledges Start-up grant from IISc, Infosys Young Researcher award, and DST-starting research grant SRG/2021/000285. The authors acknowledge funding support from the Ministry of Human Resource Development (MHRD) through NIEIN project, from Ministry of Electronics and Information Technology (MeitY) and Department of Science and Technology (DST) through NNetRA and the Thematic Unit of Excellence for Nano Science and Technology project from DST Nano Mission. A.R. acknowledges SERB-ITS Travel Grant for providing travel support. The authors thank Prof. Beatriz Noheda (RUG Groningen), Mart Salverda (RUG Groningen) and Ruben Hamming-Green (RUG Groningen) for their valuable support in ex-situ performing electrical measurements prior to making TEM samples.

**References**

[1]  D. R. Chialvo, *Nat. Phys.* **2010**, *6*, 744.

[2]  V. Mnih, K. Kavukcuoglu, D. Silver, A. A. Rusu, J. Veness, M. G. Bellemare, A. Graves, M. Riedmiller, A. K. Fidjeland, G. Ostrovski, S. Petersen, C. Beattie, A. Sadik, I. Antonoglou, H. King, D. Kumaran, D. Wierstra, S. Legg, D. Hassabis, *Nature* **2015**, *518*, 529.

[3]  W. Schultz, P. Dayan, P. R. Montague, *Science (80-. ).* **1997**, *275*, 1593.

[4]  K.-H. Kim, S. Gaba, D. Wheeler, J. M. Cruz-Albrecht, T. Hussain, N. Srinivasa, W. Lu, *Nano Lett.* **2012**, *12*, 389.

[5]  I. Sourikopoulos, S. Hedayat, C. Loyez, F. Danneville, V. Hoel, E. Mercier, A. Cappy, *Front. Neurosci.* **2017**, *11*, DOI 10.3389/fnins.2017.00123.

[6]  P. M. Sheridan, F. Cai, C. Du, W. Ma, Z. Zhang, W. D. Lu, *Nat. Nanotechnol.* **2017**, *12*, 784.

[7]  C. Li, M. Hu, Y. Li, H. Jiang, N. Ge, E. Montgomery, J. Zhang, W. Song, N. Dávila, C. E. Graves, Z. Li, J. P. Strachan, P. Lin, Z. Wang, M. Barnell, Q. Wu, R. S. Williams, J. J. Yang, Q. Xia, *Nat. Electron.* **2018**, *1*, 52.

[8]  T. Chavan, S. Dutta, N. R. Mohapatra, U. Ganguly, *IEEE Trans. Electron Devices* **2020**, *67*, 2614.

[9]  P. Bak, C. Tang, K. Wiesenfeld, *Phys. Rev. A* **1988**, *38*, 364.

[10] J. M. Beggs, D. Plenz, *J. Neurosci.* **2003**, *23*, 11167.

[11] A. Sornette, D. Sornette, *Europhys. Lett.* **1989**, *9*, 197.

[12] J. S. Urbach, R. C. Madison, J. T. Markert, *Phys. Rev. Lett.* **1995**, *75*, 276.

[13]     P. J. Cote, L. V. Meisel, *Phys. Rev. Lett.* **1991**, *67*, 1334.

[14]     B. Casals, G. F. Nataf, E. K. H. Salje, *Nat. Commun.* **2021**, *12*, 1.

[15]     C. R. Myers, J. P. Sethna, *Phys. Rev. B* **1993**, *47*, 11171.

[16]     J. P. Sethna, K. A. Dahmen, C. R. Myers, *Nature* **2001**, *410*, 242.

[17]     T. W. Boonstra, B. J. He, A. Daffertshofer, *Front. Physiol.* **2013**, *4*, DOI 10.3389/fphys.2013.00079.

[18]     J. M. Palva, A. Zhigalov, J. Hirvonen, O. Korhonen, K. Linkenkaer-Hansen, S. Palva, *Proc. Natl. Acad. Sci.* **2013**, *110*, 3585.

[19]     J. B. Mallinson, S. Shirai, S. K. Acharya, S. K. Bose, E. Galli, S. A. Brown, *Sci. Adv.* **2019**, *5*, DOI 10.1126/sciadv.aaw8438.

[20]     D. Marković, C. Gros, *Phys. Rep.* **2014**, *536*, 41.

[21]     W. L. Shew, H. Yang, T. Petermann, R. Roy, D. Plenz, *J. Neurosci.* **2009**, *29*, 15595.

[22]     E. Bullmore, O. Sporns, *Nat. Rev. Neurosci.* **2012**, *13*, 336.

[23]     J. Hochstetter, R. Zhu, A. Loeffler, A. Diaz-Alvarez, T. Nakayama, Z. Kuncic, *Nat. Commun.* **2021**, *12*, DOI 10.1038/s41467-021-24260-z.

[24]     S. Shirai, S. K. Acharya, S. K. Bose, J. B. Mallinson, E. Galli, M. D. Pike, M. D. Arnold, S. A. Brown, *Netw. Neurosci.* **2020**, *4*, 432.

[25]     M. C. Lemme, D. Akinwande, C. Huyghebaert, C. Stampfer, *Nat. Commun.* **2022**, *13*, 2.

[26]     A. Rao, S. Raghavan, *J. Mater. Chem. C* **2022**, *10*, 10412.

[27]     J. Miao, B. Wang, C. V. Thompson, *Phys. Rev. Mater.* **2020**, *4*, DOI 10.1103/PhysRevMaterials.4.043608.

[28]     B. J. Kooi, *Phys. Rev. B* **2004**, *70*, 224108.

[29]     S. K. Acharya, E. Galli, J. B. Mallinson, S. K. Bose, F. Wagner, Z. E. Heywood, P. J. Bones, M. D. Arnold, S. A. Brown, *ACS Appl. Mater. Interfaces* **2021**, *13*, 52861.

[30]     A. J. Mandell, K. A. Selz, *J. Stat. Phys.* **1993**, *70*, 355.

[31]     M. Garofalo, T. Nieus, P. Massobrio, S. Martinoia, *PLoS One* **2009**, *4*, e6482.

[32]     P. Massobrio, V. Pasquale, S. Martinoia, *Sci. Rep.* **2015**, *5*, 10578.

[33]     M. C. Kuntz, J. P. Sethna, *Phys. Rev. B - Condens. Matter Mater. Phys.* **2000**, *62*, 11699.

[34]     Z. Heywood, J. Mallinson, E. Galli, S. Acharya, S. Bose, M. Arnold, P. Bones, S. Brown, *Neuromorphic Comput. Eng.* **2022**, *2*, 024009.

[35]     N. Marshall, N. M. Timme, N. Bennett, M. Ripp, E. Lautzenhiser, J. M. Beggs, *Front. Physiol.* **2016**, *7*, 1.

[36]     N. Friedman, S. Ito, B. A. W. Brinkman, M. Shimono, R. E. L. DeVille, K. A. Dahmen, J. M. Beggs, T. C. Butler, *Phys. Rev. Lett.* **2012**, *108*, 208102.

[37]	N. Sheremetyeva, D. Niedzielski, D. Tristant, L. Liang, L. E. Kerstetter, S. E. Mohney, V. Meunier, *2D Mater.* **2021**, *8*, 025031.

[38]	H. Häkkinen, M. Moseler, O. Kostko, N. Morgner, M. A. Hoffmann, B. v. Issendorff, *Phys. Rev. Lett.* **2004**, *93*, 093401.

[39]	F. Iyikanat, H. Sahin, R. T. Senger, F. M. Peeters, *APL Mater.* **2014**, *2*, DOI 10.1063/1.4893543.

[40]	M. J. Lee, S. H. Kim, S. Lee, C. Yoon, K. A. Min, H. Choi, S. Hong, S. Lee, J. G. Park, J. P. Ahn, B. H. Park, *NPG Asia Mater.* **2020**, *12*, DOI 10.1038/s41427-020-00272-x.

[41]	G. Zhang, C. Liu, L. Guo, R. Liu, L. Miao, F. Dang, *Adv. Energy Mater.* **2022**, DOI 10.1002/aenm.202200791.

[42]	S. Menzel, U. Böttger, M. Wimmer, M. Salinga, *Adv. Funct. Mater.* **2015**, *25*, 6306.

[43]	G. Milano, G. Pedretti, K. Montano, S. Ricci, S. Hashemkhani, L. Boarino, D. Ielmini, C. Ricciardi, *Nat. Mater.* **2022**, *21*, 195.

[44]	A. Loeffler, R. Zhu, J. Hochstetter, A. Diaz-Alvarez, T. Nakayama, J. M. Shine, Z. Kuncic, *Neuromorphic Comput. Eng.* **2021**, *1*, 014003.

# Supporting Information

## Self-assembled neuromorphic networks at self-organized criticality in Ag-hBN platform

*Ankit Rao[1], Sooraj Sanjay[1], Majid Ahmadi[2], Anirudh Venugopalrao[1], Navakanta Bhat,[1] Bart Kooi[2,3], Srinivasan Raghavan[1], Pavan Nukala[1]\**

## Affiliations

[1]Centre for Nano Science and Engineering, Indian Institute of Science, Bengaluru 560012, [2]Zernike Institute for Advanced Materials, University of Groningen, Groningen, the Netherlands, [3]CogniGron center, University of Groningen, Groningen, 9747 AG, The Netherlands. Corresponding Author. * E-mail: pnukala@iisc.ac.in

**This file includes:**

Figures S1 to S7

Table S1

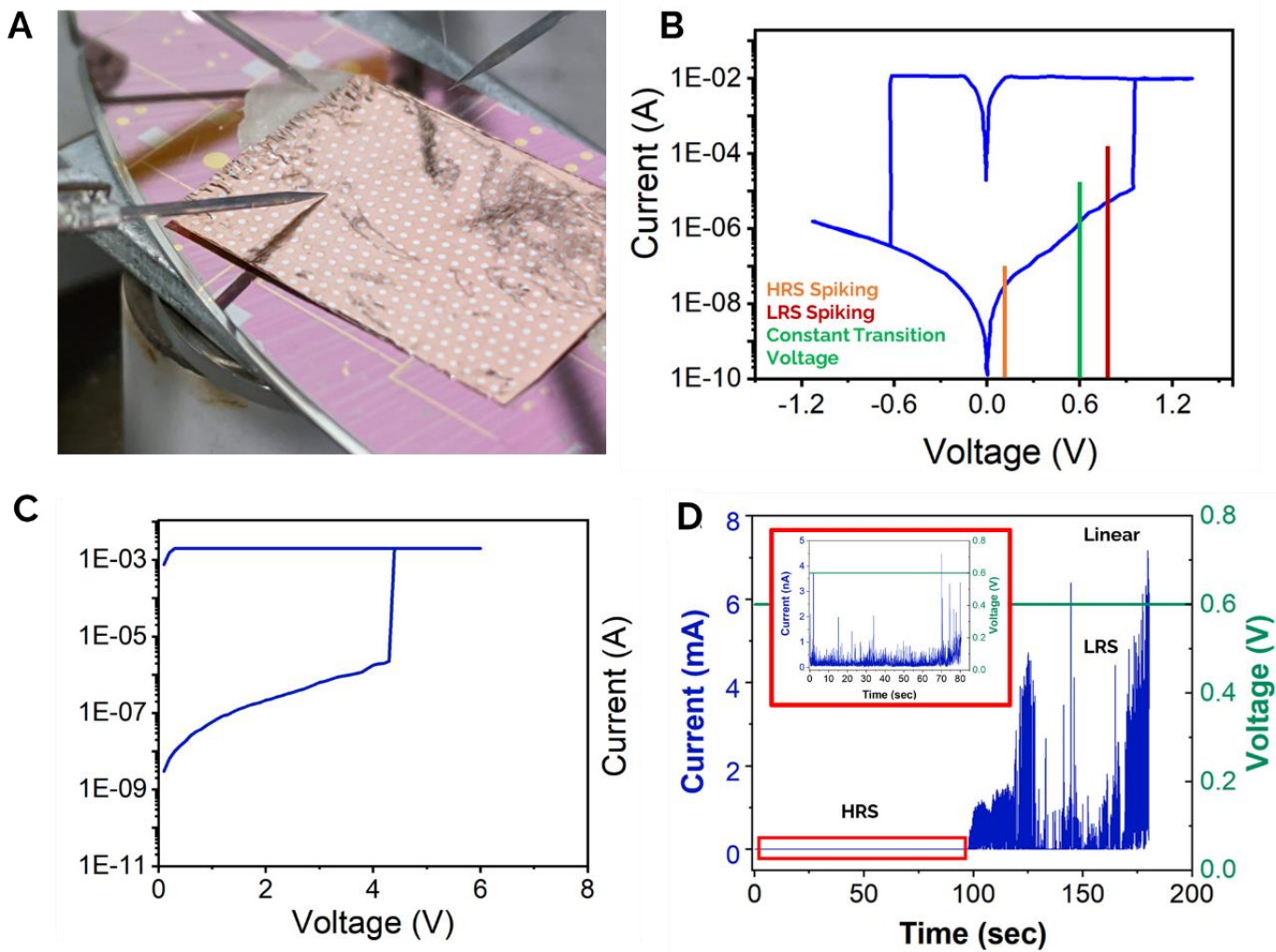

**Figure S1.** A) Large Area Ag-hBN-Cu device. B) Resistive switching behavior observed in the device at SET voltage (0.9V) and RESET voltage (0.6V). C) Electroforming process for the Ag-hBN device with forming voltage ($V_f$) of ~4V. D) Current-time plot in linear scale at a constant voltage of 0.6V. The temporal behavior of device exhibiting transition from HRS to LRS due to transport of ions under constant electric field. (Inset) The HRS pattern observed scaled to the current levels of the device in nA.

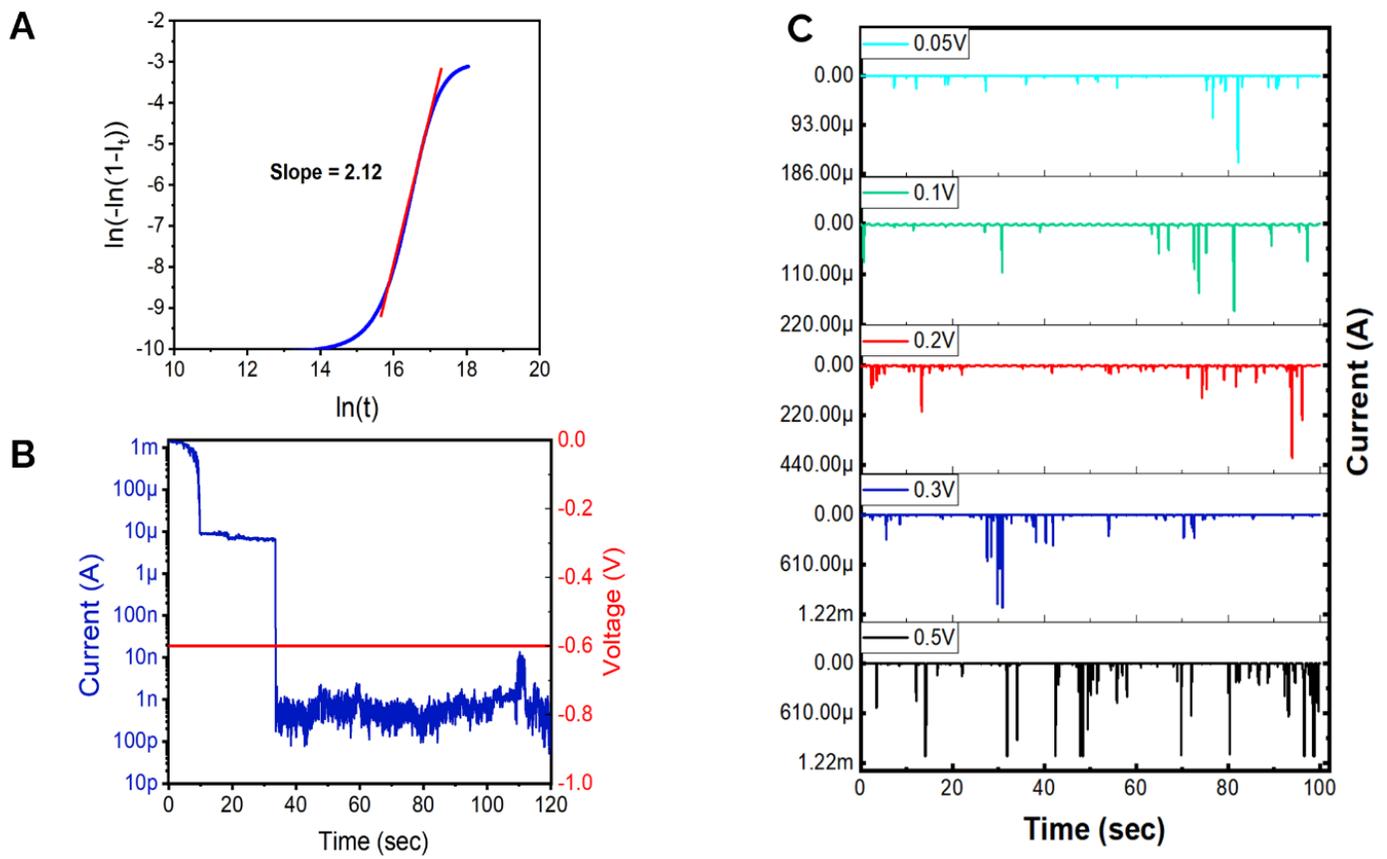

**Figure S2.** A) Avrami exponent determination from the ln(-ln(1-x) vs ln(t) curve. B) Time series plot of current under constant negative bias (-0.6V) resulting in filamentary breakage and reversal phenomenon. C) Rate coding. Controlling spiking rate as a function of input voltage. Spiking rate can be used to encode information for a specific input voltage.

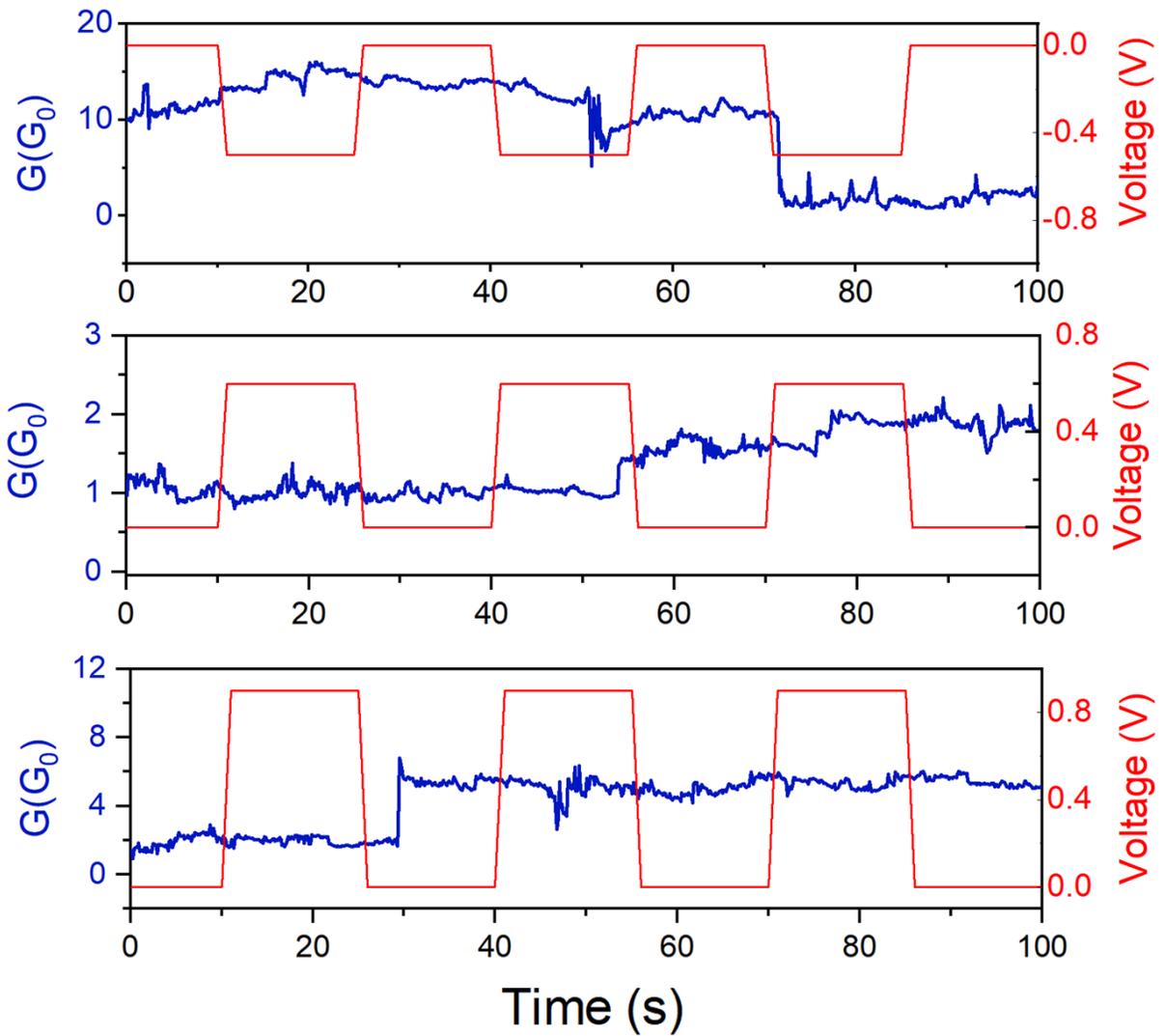

**Figure S3.** Diffusion based self-stabilizing behavior under application of long-range pulsed stimulus. (Top Panel) Under negative voltage stimulus (-0.6V), when the system is at higher conductance levels, the filaments tend to break and the return to lower conductance levels. (Middle Panel) Pulsing at a voltage (0.6V) lower than the SET voltage allows the system to stay in a stable configuration with small temporal change in conductance. (Bottom Panel) Under pulsing at voltage (0.8V) closer to SET point, the system conductance increases due to diffusion under application of higher fields.

**Long Range Pulsed Stimulus**

When negative voltage pulses in LRS state are applied (-0.6V), the system tends to switch from a high conductance level and the return to low conductance level as the filaments tend to break. When the device is in HRS state, the pulsing is applied at a voltage (0.6V) lower than the SET voltage which allows the system to stay in a stable configuration as small temporal changes in conductance are observed. Under pulsing at voltage (0.8V) closer to SET point, the system conductance increases due to diffusion under application of higher fields.

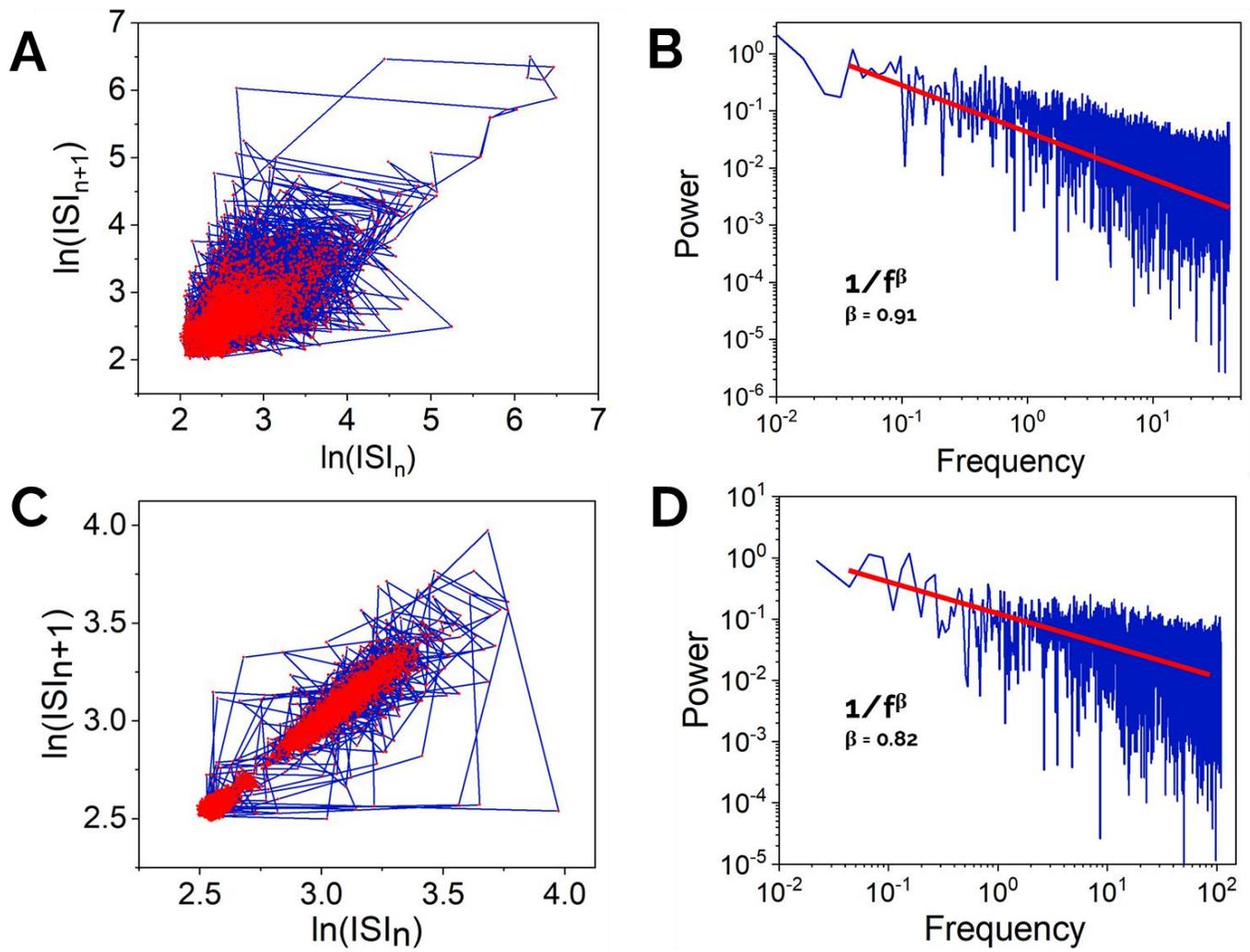

**Figure S4.** (A and C) Correlation between successive inter-spike intervals (ISI$_n$) for HRS and LRS cases. (B and D) Power Distribution (A$^2$/Hz) as a function of frequency (Hz) plot for determination of 1/f noise and β exponent value for HRS and LRS.

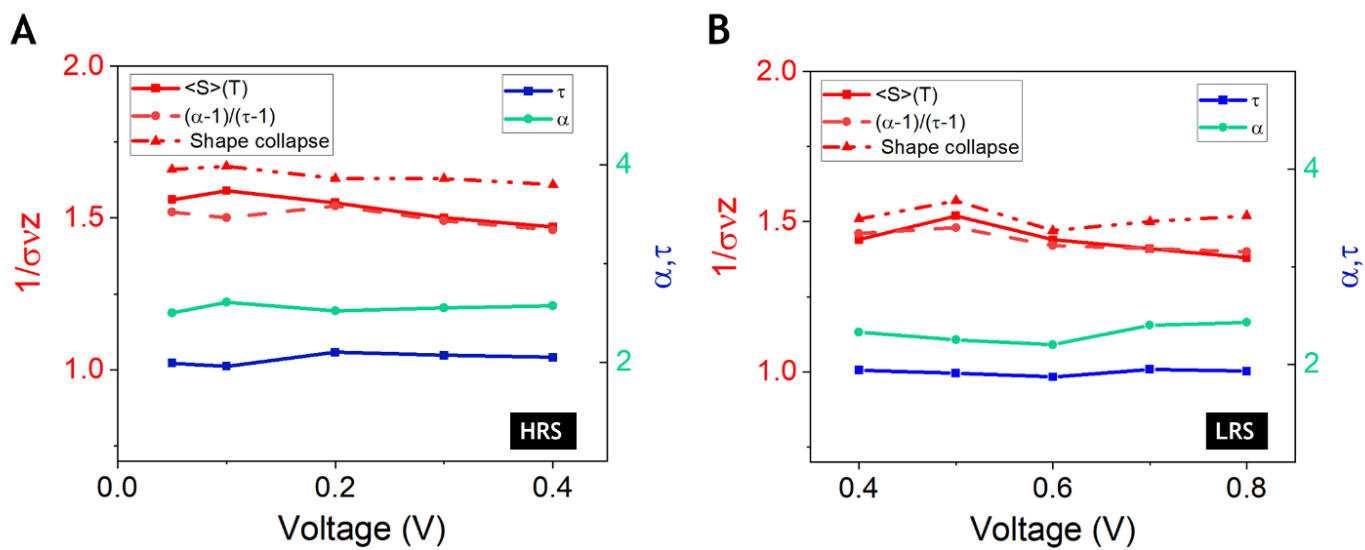

**Figure S5.** Variation of estimations of $1/\sigma\upsilon z$ as a function of voltage from $<S>(T)$ vs T, $(\alpha-1)/(\tau-1)$ and shape collapse analysis. Trend of critical exponents ($\alpha$, $\tau$) as a function of voltage for HRS (A) and LRS (B).

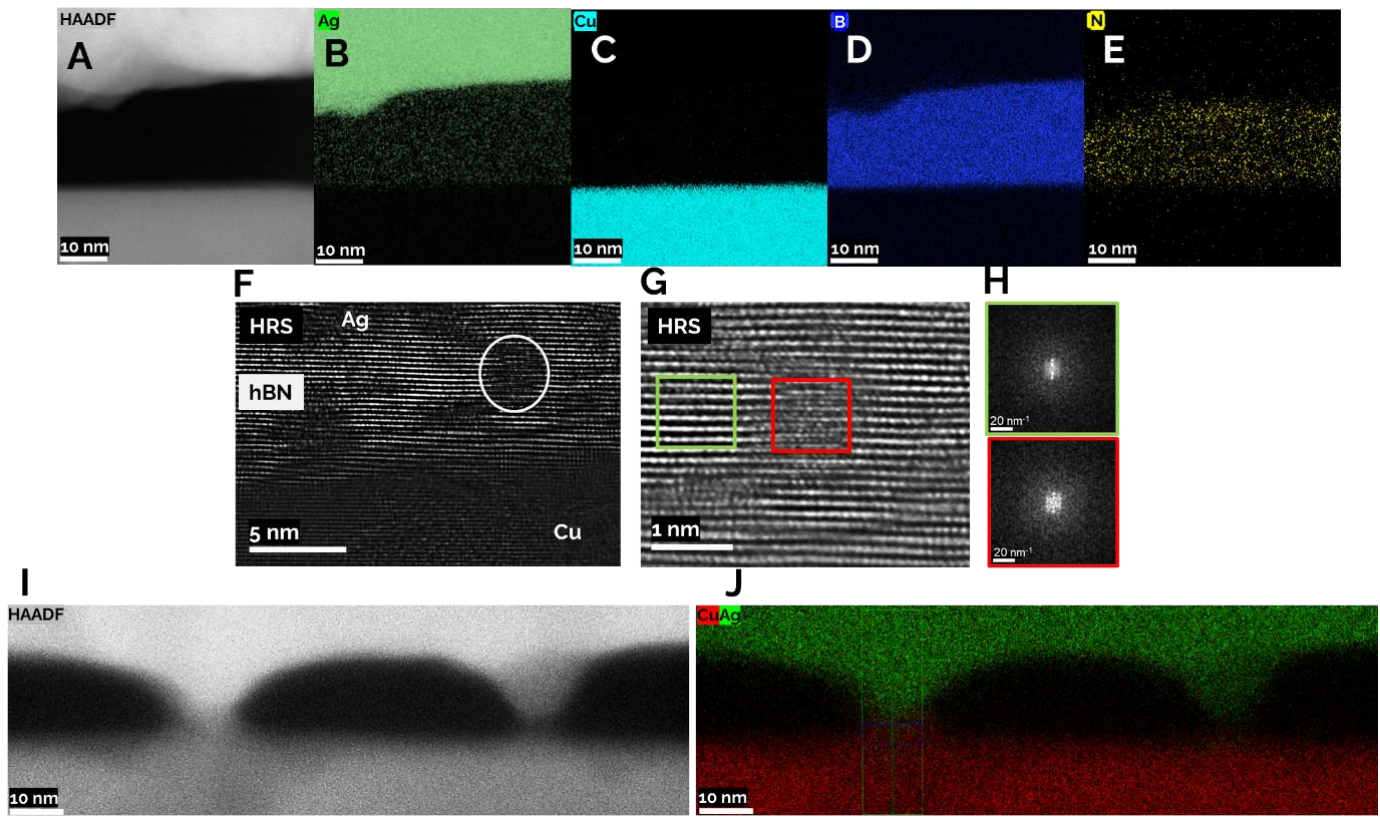

**Figure S6.** (A to E) HAADF-STEM EDS mapping of Ag-hBN-Cu stack in the HRS state. F) HRTEM image of the cross section from same sample after 10 switching cycles and rested in HRS state. G) Magnified area from F) (white region) which suggests the presence of an intercalated planar cluster of Ag between two hBN layers. H) The Fast Fourier Transform (FFT) of the regions pristine (green) and defective (red) from G) confirms change in the crystallinity of the structure due to Ag intercalation. I, J) HAADF-STEM (I) EDS (J) mapping of the Ag-hBN-Cu in the LRS state showing formation of Ag filaments.

**Table S1. Critical Exponents**

|  | ΔG | IEI | τ | α | 1/συz |
|---|---|---|---|---|---|
| **HRS** | 2.59±0.05 | 1.54±0.05 | 1.98±0.1 | 2.61±0.1 | 1.59±0.03 |
| **LRS** | 2.87±0.05 | 1.72±0.05 | 1.93±0.1 | 2.43±0.1 | 1.38±0.03 |
| **Nanoparticle Network** | 2.59 | 1.93 | 2.05±0.1 | 2.66±0.1 | 1.55±0.03 |
| **Nanowire Network** | - | - | 2±0.1 | 2.3±0.1 | 1.3±0.03 |
| **Rat cortices** | - | - | 1.74 | 1.96 | 1.22 |
| **Neuronal cortices** | - | - | 1.69 | 1.92 | 1.56 |

(**\*Error Analysis** - This standard deviation is used as a measure of the error of the exponent of the original fit, and is estimated from the routines presented in Ref 36)

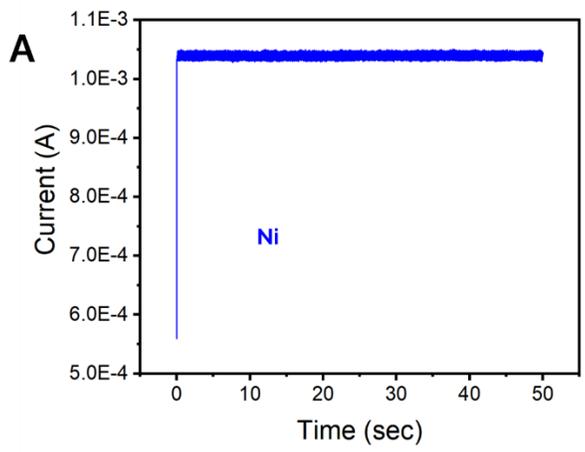 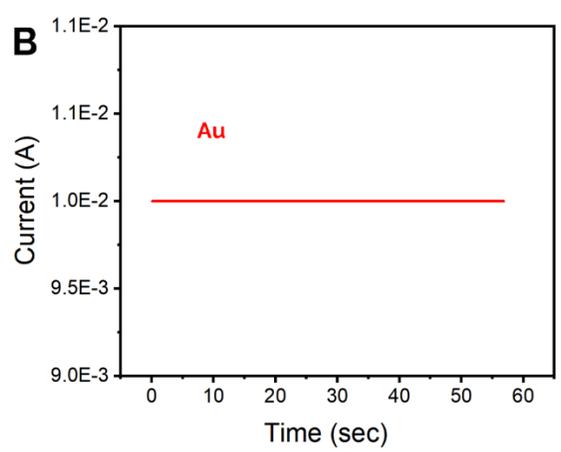

**Figure S7.** A) Ni and B) Au as the top electrode. No spiking behavior is observed with Ni and Au as the top electrode.